  \documentstyle[11pt,aaspp4]{article}  

\slugcomment{ApJ, submitted, September 16, 1999}

\lefthead{Fenimore \& Ramirez-Ruiz} \righthead{GRB by
Internal Shocks and Deceleration} 
\begin{document} \title{Gamma-Ray Bursts
as Internal Shocks Caused by Deceleration} \author{E.~E.~Fenimore$^{1}$,
and E.~Ramirez-Ruiz$^{1,2}$} \affil{$^1$MS D436, Los Alamos National
Laboratory, Los Alamos, NM 87545} \affil{$^2$Facultad de Ciencias,
Universidad Nacional Aut\'onoma de M\'exico, Distrito Federal, M\'exico
04510}
\begin{abstract} Gamma-ray bursts (GRBs) have been thought to
originate from internal shocks that occur about $10^{15}$ cm from a
central site. The shells responsible for these shocks merge together and
undergo an external shock at $\sim 10^{17}$ cm, producing the afterglows.  
We include deceleration in our model of internal shocks and find that, for
values of the Lorentz factor greater than $10^3$, deceleration is an
effective catalyst for converting the bulk motion energy into radiation
during the GRB phase.  Substantial internal energy occurs because other
shells run into the back of the first shell which has decelerated and
because the first shell must energize the interstellar medium. Whereas
internal shocks without deceleration are about 25\% efficient, we can
convert up to 85\% of the bulk motion energy during the GRB phase.  We
demonstrate that the resulting time history can have three components. The
first is due to internal shocks, but not those that involve the first
shell. This component produces narrow peaks throughout the time history.
The second is due to internal shocks involving the first shell, and it
produces progressively wider and wider peaks but they tend to be hidden in
a slowly varying background in the event. The third component is from
energizing the interstellar medium. It is very smooth and may contribute
mostly to a lower energy bandpass than the BATSE experiment.  There have
been claims of upper limits on the possible Lorentz factor because the
deceleration must occur at greater radii than the internal shocks to avoid
making progressively wider peaks. We do not find this to be the case, and
the Lorentz factor can be much larger.

\end{abstract}

\keywords{gamma-rays: bursts}

\def\domega{{\rm d}$\Omega$}
\def\lognp{Log~$N$-Log~$P$}
\def\Mesz{M\'esz\'aros}
\def\lap{\hbox{{\lower -2.5pt\hbox{$<$}}\hskip -8pt\raise-2.5pt\hbox{$\sim$}}}
\def\gap{\hbox{{\lower -2.5pt\hbox{$>$}}\hskip -8pt\raise-2.5pt\hbox{$\sim$}}}
\def\ul#1{$\underline{\smash{\vphantom{y}\hbox{#1}}}$}
\def\INITIALCON{{4\pi E_{54} \over \rho_0{\rm d}\Omega}}
\section{INTRODUCTION}
Gamma-ray bursts (GRBs) are characterized by chaotic time histories
which are often followed by x-ray, optical, and
radio afterglows (\cite{metzger97,cos97,frail97}).  The optical
afterglows have
shown redshifted absorption
features which firmly established a cosmological distance scale for the
events. The distance implies that GRBs emit the order of 
$10^{52}$ to
$10^{54}$ erg (assuming  isotropic emission).
GRBs are also often characterized by
emission up
to 100 MeV  with occasional reports of emission up to 10 Gev
(\cite{hurley94}).
Given the photon density implied by the cosmological distances,
photons above $\sim$ 1 MeV would be destroyed by photon-photon attenuation
if the emission is isotropic in our rest frame.  Large relativistic bulk
motion (Lorentz factors of $\gap$ 100) allows for a much larger emitting
surface combined with relativistic beaming that reduces the photon-photon
attenuation (\cite{fen93}). 

 The high Lorentz factor plays a crucial role in virtually all models of
GRBs.  Originally, the prime suspect for the source of $10^{52}$ erg was
a neutron star-neutron star merger (\cite{pac86}).  However, such mergers
were thought to
occur on timescales (a few millisec) much shorter than GRB timescales (up
to $10^3$ sec).  \cite{mr93n} suggested that a
relativistic shell would be formed by the initial release of energy.  The
shell could emit for a long time ($10^7$ sec). If the shell is mostly
moving directly at the observer, the shell stays close to the photons it
emits such that they
all arrive at a detector over a short period of time. If the shell moves
with
velocity $v$, then photons emitted over a duration $\Delta t$ arrive at
the
detector compressed into a duration of only $(c-v)\Delta t/c \approx
\Delta t/(2\Gamma^2)$ where $\Gamma$ is the bulk Lorentz factor =
$(1-\beta^2)^{-1/2}$ and $\beta=v/c$.  The shell would emit due to the
formation of an ``external'' shock when the shell decelerates by sweeping
up the interstellar medium (ISM). In this explanation, density variations
in the ISM cause the
observed time structure. The
deceleration is expected to occur at
\begin{equation}
R_{\rm dec} = 5 (\rho E_{0})^{1/3}\Gamma_0^{-2/3} {\rm cm}
\label{RDEC}
\end{equation}
where $\rho$ is the ambient density (in cm$^{-3}$), $E_{0}$ is total
energy (in erg) generated by 
the central site, and
$\Gamma_0$ is
the initial Lorentz factor (\cite{rm92}). For typical values such as
$\rho=1$ cm$^{-3}$, $E_0=10^{53}$ erg, and $\Gamma_0=100$, the
deceleration
occurs at about $10^{17}$ cm. The initial Lorentz factor is set by the
baryon loading, that is 
\begin{equation}
E_0 = \Gamma_0 m_0 c^2
\label{E0}
\end{equation}
where $m_0$ is the mass of the shell, presumably carried by the baryons.

  An alternative explanation is that the central site produced a series of
shells.  Collisions between shells produce the gamma rays through internal
shocks (\cite{rm94}). The faster shells catch up with the slower shells.
The collision radius is roughly
\begin{equation}
R_{\rm col} = c \Gamma^2 \Delta T
\label{RCOL}
\end{equation}
where $\Delta T$ is a typical time of variation in a GRB
(\cite{rm94}). For
typical values such as $\Delta T$ = 0.1 to 1.0 s, $R_{\rm col}$ is about
$10^{15}$ cm.  The observed duration of the GRB is set by
the duration of the activity at the central site. 

There are a series of arguments that indicate that the gamma-ray phase is
not caused by external shocks.  These arguments are all related to the
fact that the size of the shell at the deceleration radius is much larger
than a causally connected region on the shell.  For example, precursors
and gaps
require large causally disconnected regions to coordinate their activity
(\cite{fmn96}). The observed variability implies that only a small
fraction
of the shell emits (\cite{fmn96etal,sp97}), and
the average profile of many GRBs is inconsistent with that expected from a
shell (\cite{fen99}).  Finally,
the constancy of the pulse
width throughout the bursts indicates that we are not seeing a range of
angles on a shell and that the shell is not decelerating
(\cite{enricoa}). A single decelerating shell would produce pulses that
get
progressively wider.

However, external shocks have been very successful is explaining the
x-ray, optical, and radio afterglows (see review by, e.g.,
\cite{piranrev}). These afterglows usually show power
law decays and spectral variability expected from a decelerating shell.
Thus, a general picture has formed where the central site produces
multiple shells for tens of seconds. These shells collide, producing the
gamma-ray phase by internal shocks and then merge into a single shell
which interacts with the ISM to produce the afterglows
(\cite{sp97}). 
One-dimensional hydrodynamical calculations have reproduced many of
the features of the spectral evolution (\cite{panm98}).
There have been several detailed calculations of what is expected from the
internal shocks. \cite{moch}, \cite{ksp97n}, and \cite{dm98} used
 Monte Carlo calculations
of
internal shocks where the $\Gamma$, mass, time, and thickness of multiple
shells are picked randomly to demonstrate that internal shocks can
produce the variability seen in GRBs.
  By following their trajectories, it is
determined when they collide.  The duration of the resulting simulated
GRBs is effectively the duration of the activity at the central site.  The
rise of each pulse depends on the time for a reverse shock to cross
the shell and the fall depends on the curvature of the shell at the radius
of interaction (\cite{ksp97n}).  The radius of interaction depends on the
Lorentz factors
and the amount of time between the production of the shells at the central
site.  Indeed, the resulting time histories  bear some similarities to the
observed
bursts.  

  The strong optical emission discovery by the Robotic Optical Transient
Search Experiment, ROTSE, (\cite{carl}) as served as an excellent test
case for the external shock model. \cite{sp99bn}  and \cite{mr99} fit
 forward and reverse
external shocks and had excellent agreement with the time of the optical
peak, the rise and fall times, the overall magnitude, and the break in
the decay phase.

\section{The Role of  {\bf $\Gamma$}}

The Lorentz factor is not well determined observationally.  The lack of
apparent photon-photon attenuation  up to $\sim$ 100 MeV implies only a
{\it lower} limit of $\sim 100$ (\cite{fen93}). 
The $\Gamma$ determined by \cite{sp99bn} for GRB990123 depended
on some parameters adopted from \cite{wijers99n} for GRB970508, but gave a
similar low value of $\Gamma=200$. However, there have been recent reports
of possible TeV emission from GRBs (\cite{leonor99}) implying that $\Gamma$
may be much larger.  Also, if $\Gamma$ is small,  the efficiency of
internal shocks is small, the order of 10\%.

 The first shell starts to decelerate when it has swept up $\sim
\Gamma^{-1}$
of its initial mass.  Thus, larger $\Gamma$ means the deceleration will 
occur at a smaller radius.  But, a larger $\Gamma$
means that the multiple shells will collide at a larger radius.
Combining equations (\ref{RDEC}) and (\ref{RCOL}),
internal shocks will occur at about the same radii as the deceleration
if $\Gamma \gap 10^{-10}(\rho E_0)^{1/4}$.
From the few observed redshifts, we know that GRBs
have a rather large range of fluences, with typical values between
$10^{52}$ and $10^{54}$ erg for isotropic emission.  We have little direct
knowledge of the
ambient density in the vicinity of a GRB, but it is reasonable to assume
values of $\rho$ between 0.1 and 10.0, so $\rho E_0$ varies from
$10^{51}$ to $10^{55}$ erg cm$^{-3}$.
For values of $\Gamma$ of 
 $\sim 500$ to $3000$, the
internal
shocks will occur
about the same place as the deceleration.

Once the first shell decelerates
(making an external shock), the
rest of the shells will rapidly catch up to it resulting in rather 
efficient internal
shocks. In previous models (e.g.,
\cite{ksp97etal}), $\Gamma$ was about
$10^2$ to $10^4$ but deceleration was not included.  It was assumed that
the internal shocks would form at small radii and later the merged shell
would suffer deceleration and an external shock.  Perhaps a few straggler
shells would catch up to the first shell after it decelerated and
rejuvenate it during the afterglow phase (\cite{panmr98}), but most of the
gamma-rays were
assumed to form at small
radii relative to the external shock.

We propose that the typical Lorentz factor is large enough such that the
first shell decelerates before all of the multiple shells have a chance to
collide.  The deceleration is very rapid once it starts,
effectively equivalent to slamming on the breaks.  The rest of the shells
catch up and collide with it. Since the efficiency of converting
bulk motion to radiation in an internal shock depends on the difference of
the colliding $\Gamma$'s, the fact that the first shell is decelerating
implies that the efficiency will be higher than in previous models.  
Thus, the collisions are internal shocks but the place and efficiency of
the many of collision are caused by deceleration.

\section{Ingredients for a Model}

  In the internal shock model, multiple shells are generated by an
unspecified process at a central site. 
The parameters of our model will be similar to those
of \cite{moch}, \cite{ksp97netal} and \cite{dm98},
including the time the $i$-th shell was
generated ($t_{0i}$), the initial width of the shell ($l_i$), the minimum
and maximum initial Lorentz factor ($\Gamma_{\rm min}, \Gamma_{\rm max}$),
and the initial energy ($E_i$).
\cite{ksp97netal} allowed for selecting the initial mass, energy, or
density.  However, all three gave similar results and we
will restrict ourselves to selecting the energy.
The initial  $m_i$ is then found
from equation
(\ref{E0}). Since \cite{ksp97netal} presented unitless intensities, it was
unnecessary for them to specify $E_i$.  The bulk energy is necessary to
set the deceleration.  
The peak energy can be estimated from 
bursts with  observed redshifts.  GRB970508 had a peak luminosity,
$L$, of $\sim
3\times 10^{51}$ erg s$^{-1}$.  Other GRBs have shown extreme
redshifts
(e.g., \cite{kulkarni0123}), implying $L \sim 2 \times 10^{53}$ erg
s$^{-1}$.
We will uniformly select $E_i$ between  $E_{\rm min}$ (=$10^{49}$ erg
s$^{-1}$) and $E_{\rm max}$, and vary $E_{\rm max}$ from $10^{51}$ to
$10^{53.5}$ erg.

We randomly select
$t_{0i+1}-t_{0i}$ from a Poisson
distribution based on the rate of peak occurrence.  Thus, we specific
the
duration of the activity at the central site ($T_{\rm dur}$) and the
expected number of peaks ($N$) such that the actual number of peaks is
random. Since they were presenting results in unitless time,
\cite{ksp97netal} set the burst duration, shell separation,  and the
number of peaks
to be 
constants.  These differences are not important when there is no
deceleration. 
We will use parameters that are roughly
equivalent to \cite{ksp97netal}, that is, $N = 85$, $T_{\rm dur}$ = 60
s, and $l_i$ = 0.2 s.  Burst often have gaps which implies that the
activity at the central site can turn off for a while. To demonstrate the
effects of turning off the central site, we impose a gap in the activity
between $T = 20$ and $T = 33$ s.

Until they collide with the first shell or each other, the
motion of the every shell except the first is constant,
$R_i(t)=c\beta_i(t-t_{0i})$. If $\Gamma_i$ is greater than $\Gamma_j$,
the two shells will collide at time $t_{ij}$ when
$R_i(t_{ij})=R_j(t_{ij})$, which we
call the collision radius (=$R_c$), and it occurs at
\begin{equation}
t_{ij} = 2{\Gamma_i^2\Gamma_j^2 \over \Gamma_i^2-\Gamma_j^2}\Delta t_{0ij}
\label{TCOLL}
\end{equation}
where $\Delta t_{0ij} = t_{0i}-t_{0j}$. The resulting pulse arrives at a
detector at the  relative time of arrive
\begin{equation}
T_{\rm toa} = t_{ij}-R_c/c = t_{oi}+{\Gamma_j^2 \over
\Gamma_i^2-\Gamma_j^2}\Delta t_{0ij}~~.
\label{TOA}
\end{equation}
Thus, the relative time of arrival at a detector will have a close
one-to-one relationship with  the time
the shell was created (i.e., $t_{0i}$).

In order to conserve both momentum and energy when shells collide,
some of the bulk energy must be converted to internal energy which will be
radiated away. If $E_{\rm rad}$ is the generated internal energy,
conservation of energy dictates that
\begin{equation}
m_i\Gamma_i + m_j\Gamma_j = \big[m_{ij}+{E_{\rm rad}\over c^2}\big]
\Gamma_{ij}
\label{ECON}
\end{equation}
where $\Gamma_{ij}$ is the Lorentz factor of the resulting shell and
the resulting mass is $m_{ij}=m_i+m_j$.  Conservation of momentum gives:
\begin{equation}
m_i\beta_i\Gamma_i + m_j\beta_j\Gamma_j = \big[m_{ij}+
{E_{\rm rad}\over c^2}\big]\beta_{ij}\Gamma_{ij}
\label{CONMV}
\end{equation}
where, as usual, the  $\Gamma$ terms are related to the $\beta$ terms as
$\Gamma = (1-\beta^2)^{-1/2}$.
The post collision $\beta$ is
\begin{equation}
\beta_{ij} = {m_i\beta_i\Gamma_i + m_j\beta_j\Gamma_j \over
              m_i\Gamma_i + m_j\Gamma_j}
\label{POSTBETA}
\end{equation}
which has the approximate solution (\cite{ksp97etal})
\begin{equation}
\Gamma^2_{ij} = \Gamma_i\Gamma_j{m_i\Gamma_i + m_j\Gamma_j
                        \over m_i\Gamma_j + m_j\Gamma_i}~~.
\label{POSTGAMMA}
\end{equation}

The first shell is decelerated by the ISM. We use equations \ref{CONMV}\ and
\ref{ECON}\ with $\Gamma_{j} = 1$ and $m_j$ equal to the mass swept up during
the time step to determine the velocity of the first shell as a function
of time.

The colliding shells make a peak in the gamma-ray time history at
relative time $t_{ij}-R_c/c$. When two shells collide, forward and
reverse shocks traverse the shells.  If the
internal energy is promptly converted into radiation, the
merged shell emits for about the time that it takes for the reverse shock
to cross the shell  (see \cite{ksp97etal}).
The $\Gamma$ factors for the forward and reverse shocks are found from
\cite{sp95}
\begin{equation}
\Gamma_{fs} = \Gamma_{ij}\bigg[
   {1+2\Gamma_{ij}/\Gamma_i \over 2+\Gamma_{ij}/\Gamma_i}\bigg]^{1/2}
\label{GAMMAFS}
\end{equation}
and
\begin{equation}
\Gamma_{rs} = \Gamma_{ij}\bigg[
   {1+2\Gamma_{ij}/\Gamma_j \over 2+\Gamma_{ij}/\Gamma_j}\bigg]^{1/2}~~.
\label{GAMMARS}
\end{equation}

\section{Pulse Shape}

 A shell that
coasts without emitting
photons and then emits for a short period of time produces a pulse with a
rise time related to the time the shell emits and a decay dominated by
curvature effects (\cite{fmn96netal}).  In the internal shock model,
the  shell emits for the time it takes the reverse shock to 
cross the shell that is catching up, that is (\cite{ksp97etal}),
\begin{equation}
\Delta t_{\rm cross} = l_j/(\beta_j-\beta_{rs})~~.
\label{TCROSS}
\end{equation}

The time of arrival at a detector (relative to the start of the pulse) of
photons generated at angle $\theta$ from the line of sight is
\begin{equation}
T(\theta) = R_c(1-\cos\theta)/c
\label{TTHETA}
\end{equation}
(Note in our previous papers, it was more convenient to measure time from
when the shell left the central site; this is not used here because the
shell does not move at a constant speed.) At angle $\theta$,
the Doppler factor, $\Lambda$, is 
$\Gamma_{ij}(1-\beta_{ij}\cos\theta)$. 
At time $T$ in the pulse, the $\Lambda$ factor is
\begin{equation}
\Lambda(T) = {R_c +2\Gamma_{ij}^2cT \over 2\Gamma_{ij}R_c}
\label{LAMBDA}
\end{equation}
To calculate the observed pulse shape, one needs to combine the Doppler
beaming with the volume of material that can contribute at time $T$.
Following the method in \cite{chiphunt}, the resulting pulse shape is
\begin{eqnarray}
V(T) &=&0\hspace{325pt}{\rm if}\ T<0  \nonumber \\
&=&\psi \frac{(R_c+2\Gamma_{ij}^2cT)^{\alpha +3}
-R_c^{\alpha +3}}{(R_c+2\Gamma_{ij}^2cT)^{\alpha +1}}
\hspace{122pt}{\rm if}\ 0<2\Gamma_{ij}^2T<\Delta t_{\rm cross}
\label{CHIPENVEL} \\
&=&\psi \frac{(R_c+\Delta t_{\rm cross})^{\alpha +3}-
R_c^{\alpha +3}}{(R_c+2\Gamma_{ij}^2cT)^{\alpha+1}}
\hspace{147pt}
{\rm if}\ 2\Gamma_{ij}^2T>\Delta t_{\rm cross}
\nonumber
\end{eqnarray}
where $\psi $ is a constant and $T$ is measured from the start of the
pulse.

The cooling is very rapid so the internal energy generated by the
collision, $E_{\rm rad}$, is immediately turned into photons. An observer,
using a detector such as BATSE, sees the fraction that is in the BATSE
bandpass of 50 to 300 KeV, $f_{\rm BATSE}$.   Since we do not understand
exactly how the
internal energy is distributed, we cannot predict $f_{\rm BATSE}$.
However,
GRB often have a ``Band'' spectral shape with $\alpha=-1, \beta=-2.5,$ and
$E_{\rm peak} = 250$ KeV (\cite{band93}).
 If that shape is valid over the entire range of
emission, $f_{\rm BATSE}$ is $\sim 0.37$. We generate simulated time
histories as the sum of pulses with the shape from equation
(\ref{CHIPENVEL}) and integrated fluence of $f_{\rm BATSE}E_{\rm rad}$. We
generate the time history with 0.064 s samples (to mimic BATSE) and then
find the peak emitted luminosity in 0.256 s ($=L_{256}$). Ignoring
cosmological redshift effects, the BATSE
catalog value of $P_{256}$ should be related to $L_{256}$.

\section{A Typical Simulation}

To summarize our model, we have eight parameters: the duration of
the activity ($T_{\rm dur}$), the rate of explosions at the central site 
($N/T_{\rm dur}$), the range of Lorentz factors 
($\Gamma_{\rm min}, \Gamma_{\rm max}$), the range of energy release
at the central site ($E_{\rm min}, E_{\rm max}$), the ISM density
($\rho$), and the range of initial thicknesses (0 to
$l$).  To be comparable to \cite{ksp97netal}, we will set $N=85$, $T_{\rm
dur} = 60$ s, and $l$ = 0.2 s. \cite{ksp97netal} parameterized
much of their results based on $\Gamma_{\rm max}/\Gamma_{\rm min}$ since
the
overall efficiency of the conversion of bulk energy to radiation was
primarily dependent on that parameter. With deceleration, we have found
that the efficiency depends mostly on $\Gamma_{\rm max}$, so we have set
$\Gamma_{\rm min}$ to the minimum required for the high energy
emission (100), and varied $\Gamma_{\rm max}$ from $10^{2.5}$
to $10^{4.5}$.
\cite{ksp97netal} had no analog to $E_{\rm min}, E_{\rm max}$, and $\rho$.
Since we selected $E$ uniformly between $E_{\rm min}$ and $E_{\rm max}$,
$E_{\rm min}$ is not important as long as it is much less than $E_{\rm
max}$. We set $E_{\rm min}$ to $10^{49}$ erg. For $\rho$ we have used 1
cm$^{-3}$.

 Figure \ref{thfig}  is a typical simulation
$E_{\rm max} = 3.6 \times 10^{53}$ erg, $\Gamma_{\rm max} =
3.2 \times 10^4$. In Figure \ref{thfig}a, $\rho$ is zero, so
there is no deceleration. About $1.5 \times 10^{55}$ erg
(assuming isotropy) were released at
the central site in 81 shells.
The burst duration at the observer is approximately the duration of the
activity at the central site.
About 25\% of the bulk energy was received by the observer in the period
$T_{\rm dur}$.
Figure \ref{thfig}b is the same simulation (i.e., same set of random
numbers), but
includes
deceleration of the first shell in an ISM with $\rho = 1$ cm$^{-3}$. Both 
simulations appear similar because both reflect the activity of the
central engine (see eq. [\ref{TOA}]). The dotted line is the contribution
to the time history
from collisions with the first shell. It tends to add a DC level
with a few wide peaks but it raises the fraction of the
bulk energy
converted to radiation to 45\%. Some of the radiation will arrive after
$T_{\rm dur}$ because curvature will delay it.

Figure \ref{th_gamma}a gives the Lorentz factor for the first shell in
Figure \ref{thfig}b. Given
the high
value of $\Gamma_{\rm max}$, it quickly decelerates but other shells
collide with it, giving it a boost and maintaining a large $\Gamma$ for
most of the burst.  The deceleration occurs because the first shell
collides with the ISM. The resulting internal energy must also radiate
away. In Figure \ref{th_gamma}b we show the contribution to the time
history of Figure \ref{thfig} from the  internal energy from the
deceleration if it radiates in the BATSE bandpass. It tends to be smooth
and
would fill in gaps if it had a $f_{\rm BATSE}$ similar to that from the
internal shocks. We define $E_{\rm dec,dur}$ to be the internal energy
generated by the collision of the first shell with the ISM that would
arrive at the detector with $T_{\rm toa} < T_{\rm  dur}$. For the case in
Figure
\ref{thfig}, $E_{\rm dec,dur}$ is  38\% of the bulk motion energy.

\section{Efficiency of Converting Bulk Energy}

The efficiency of the conversion is an important constraint.  Although the
time histories imply that GRBs are central engines with internal shocks,
internal shocks usually do not convert most of the bulk motion into
energy (e.g., \lap 25\%, \cite{ksp97etal}).  Observationally, the
afterglows
only account for
a small percentage
of the energy so it is not clear where most of the energy goes.  The
efficiency for an individual collision can be found from the initial and
final bulk energies:
\begin{equation}
\epsilon_{ij} = 1-{m_{ij}\Gamma_{ij} \over m_{i}\Gamma_{i} +
m_{j}\Gamma_{j}}
\label{EFFIND}
\end{equation}
If there is no deceleration, the shells will collide until the remaining
shells are ordered with decreasing value of the Lorentz factors. Let $n$
be the number of remaining shells.    
The overall efficiency depends on how much energy remains in un-collided
shells:
\begin{equation}
\epsilon = 1-{\sum_{ij=0}^{ij=n}m_{ij}\Gamma_{ij} \over
\sum_{i=0}^{i=N}m_{i}\Gamma_{0i}}
\label{EFF}
\end{equation}
When deceleration occurs, $n$ is 1, another
reason why our model will give higher efficiency than previous models.

To study the effects of deceleration we have generated sets of 128
bursts,
under a variety of conditions.  Figure \ref{eff_l} 
shows the average efficiency as a function of $\Gamma_{\rm max}$. The
curves labeled ``No Deceleration'' is effectively the same result as
\cite{ksp97netal}. The curves labeled ``Deceleration, IS'' includes an ISM
with $\rho = 1$ cm$^{-3}$. We ran models for a range of $E_{\rm max}$
(maximum energy per shell) and interpolated the results to find the
efficiency at three values of the peak $L_{256}$ in the time histories: 
$3 \times 10^{50}, 3 \times 10^{51}$, and $3 \times 10^{52}$ erg s$^{-1}$
(the solid, dotted, and dashed lines, respectively).  Figure
\ref{radius_l} shows the corresponding average radii 
for the internal shocks that produces pulses that arrive with $T_{\rm toa}
< T_{\rm dur}$, that is, during the GRB phase.  (Including all internal
shocks would produce a misleading result when there is little deceleration
because a few stragglers would finally collide at radii orders of
magnitude
larger.)  The curves labeled ``ES'' are
the average radii at which the Lorentz factor
of the first shell is reduced by half. Once the first shell starts to
decelerate, a fair number of the shells collide with it, raising the
average amount of the bulk energy which is released in internal shocks.
These collisions are more efficient since there is a greater disparity
between the $\Gamma$ factors. For large vales of $\Gamma_{\rm max}$, the
efficiency rises to 40\%.

The curves labeled ``Deceleration, IS'' in Figure \ref{eff_l} are based
on the ratio of the internal energy generated by collisions between shells
(including the first shell) to the total generated bulk motion energy. It
does not include the bulk motion energy lost to energizing the ISM.
Eventually, all of the bulk motion energy is lost to the ISM. In previous
models it was assumed that this was far from where the internal shocks
occur.  In the curves labeled ``Deceleration, IS+ES'', we include $E_{\rm
dec,rad}$ in the efficiency, that is, the internal energy from collision
with the ISM whose photons would start to arrive during the burst.  For
large
values of $\Gamma_{\rm max}$, nearly 85\% of the bulk motion energy is
lost during the GRB phase.

\section{DISCUSSION}

  Internal shocks are capable of producing the variability that is the
signature of GRBs (\cite{ksp97etal}). However, it has been believed that
internal shocks are inefficient, converting only $\lap25$\% of the bulk
motion energy into radiation.  Since the afterglows only account for a few
percent of the radiated energy, it has been unclear where most of the
energy goes. 

In this paper, for the first time, deceleration of the first shell is
included in an internal shock model. For $\Gamma_{\rm max} \gap 10^3$,
there are two ways that deceleration is an important catalyst for
converting bulk motion into
radiation.  First, the deceleration occurs because the bulk motion must
energize the ISM that it runs into.  Much of the energy to energize the
ISM goes into internal energy. This is rather effective because the bulk
motion energy of the first shell is $\Gamma_iM_ic^2$ where the mass grows
as other shells run into the first shell. If deceleration causes the
$\Gamma$ of the first
shell to drop by 50\%, nearly 50\% of the bulk motion energy will be used.
Second, the rapid deceleration causes shells to plow into the back of the
first shell. The efficiency for converting bulk motion (for equal mass
shells) scales as $1-(\Gamma_j/\Gamma_i)^{1/2}$ if $\Gamma_i$ is much 
larger than $\Gamma_j$, as is the case when the $j$-th shell is
decelerating.  These two effects combine to release up to 85\% of the bulk
motion energy while $T_{\rm toa}$ is less than $T_{\rm dur}$, that is,
during the GRB phase.  Although the bulk motion might be effectively
converted to internal energy, the resulting electron distribution is
difficult to predict, so we do not know how much of it will occur in the
BATSE bandpass (i.e., $f_{\rm BATSE}$ is uncertain).

Thus, one can identify three types of contributions to the time history,
each with a different character. The internal shocks that do not
involve the first shell, internal shocks involving the first shell, and
the external shock produced as the first shell decelerates.

The internal shocks that do not involve the first shell are characterized
by narrow pulses, and have nearly constant width throughout the time
history (see, for example, Fig.  \ref{thfig}a). The $T_{\rm toa}$ for
these pulses is dominated by the time the shells were produced at the
central site (following eq.[\ref{TOA}]). They occur at a similar radius
from the central site. If $\Gamma_i$ is selected randomly between a small
$\Gamma_{\rm min}$ and $\Gamma_{\rm max}$, many pulses form with a similar
Lorentz factor:  $\Gamma_{ij} \sim (\Gamma_i\Gamma_j)^{1/2} \sim
\Gamma_{\rm max}/2$. The pulse shape depends mostly on $R_c$ and
$\Gamma_{ij}$ (eq. [\ref{CHIPENVEL}]), so they are quite similar.

The internal shocks involving the first shell occur at ever increasing
radii with a generally decreasing Lorentz factor. Thus, equation
(\ref{CHIPENVEL}) produces peaks that are wider and wider (see the dotted
curve in Fig.~\ref{thfig}b). In our previous papers, we argued that the
time history could not arise from a {\it single} shell because the pulses
did not get wider and wider.  This argument is still valid and this paper
shows how {\it multiple} shells can produce narrow peaks throughout the
event in the presence of wider and wider pulses from a single shell.
Indeed, a recent analysis of 387 pulses in 28 BATSE GRBs shows that the
most intense pulses in a burst have nearly identical widths throughout the
burst, but the weak pulses show a trend to become wider as the burst
progresses (\cite{err99}).  This is precisely what is seen in simulations
when deceleration is included.

There have been claims of upper limits on the possible Lorentz factor
because the deceleration must occur at greater radii than the internal
shocks (\cite{lazz99}) to avoid making progressively wider peaks. We do
not find this to be the case, and the Lorentz factor can be much larger,
allowing more of the bulk motion energy to be released during the GRB
phase.

 The third type of contribution
arises from the external shock as the first shell energizes the ISM. This
is a smooth component with some variation as the Lorentz factor increases
due to collisions with faster shells and decreases due to deceleration (see
Fig. \ref{th_gamma}b). Previous external shock models (e.g.,
\cite{dermer99}) have assumed the shock interacts with ISM clouds that are
much smaller that the size of the shell. This was necessary to produce the
temporal variability.  We do not assume any structure in the ISM so the
contribution from this component is quite smooth.

Figure \ref{batseth} shows BATSE time histories that have the
characteristics of our simulations. Figure \ref{batseth}a is BATSE burst
2831. It has many narrow peaks throughout the time history but also
gaps  that go back to background.  The presence of gaps implies little
deceleration of the first shell because the gaps are not filled in. The
gaps would occur because the central site turns on and off.
We note that this burst is the record holder for the highest energy
photons, 18 GeV (\cite{hurley94}).

 Figure \ref{batseth}b is BATSE burst 2329, and initially, it shows narrow
peaks, but then broader peaks, on top of a slower varying level.  The
structure on the rise in burst 2329 is statistically significant. The
slower varying function and widening pulses implies substantial
deceleration of the first shell.  In other bursts (e.g., BATSE burst 130),
there are gaps where the last pulse before the gap has a slow decay,
similar to that seen in Figure \ref{thfig}b. However, these gaps can go
all the way down to the background. Such bursts seem to show the signature
of internal shocks on a decelerating first shell but not the contribution
from the energization of the ISM.  Since we do not understand the
mechanism by which the internal energy is distributed, the $f_{\rm BATSE}$
values associated with each component might be different and the emission
might appear in different bandpasses. Perhaps the $f_{\rm BATSE}$ for
energizing the ISM is small and its internal energy is radiated at lower
energy such as the x-ray excesses reported by \cite{preece94}.

Apparently, some burst involve deceleration and some do not. The Lorentz
factor required to have deceleration  depends very weakly on $\rho$
and $E_0$ (i.e., $(\rho E_0)^{1/4}$). Thus it seems more likely that
intrinsic variations in $\Gamma_{\rm max}$ might be the reason why some
bursts show more deceleration than others. 


If the prompt emission is caused by the first shell, as suggested by the
analysis of GRB990123 (\cite{sp97}), we would expect events with a slowly
varying component to be more likely to slow prompt, bright optical
emission or early afterglows.

In summary, it is possible to convert a large fraction ($\sim 85$\%) of
the bulk motion energy into radiation during the gamma-ray burst phase
with internal shocks if deceleration of the first shell is account for and
the Lorentz factor is $\gap 10^3$.

{\it acknowledgment:} The authors gratefully acknowledge useful conversations
with Re'em Sari.

%

\clearpage
%

%
%
\figcaption[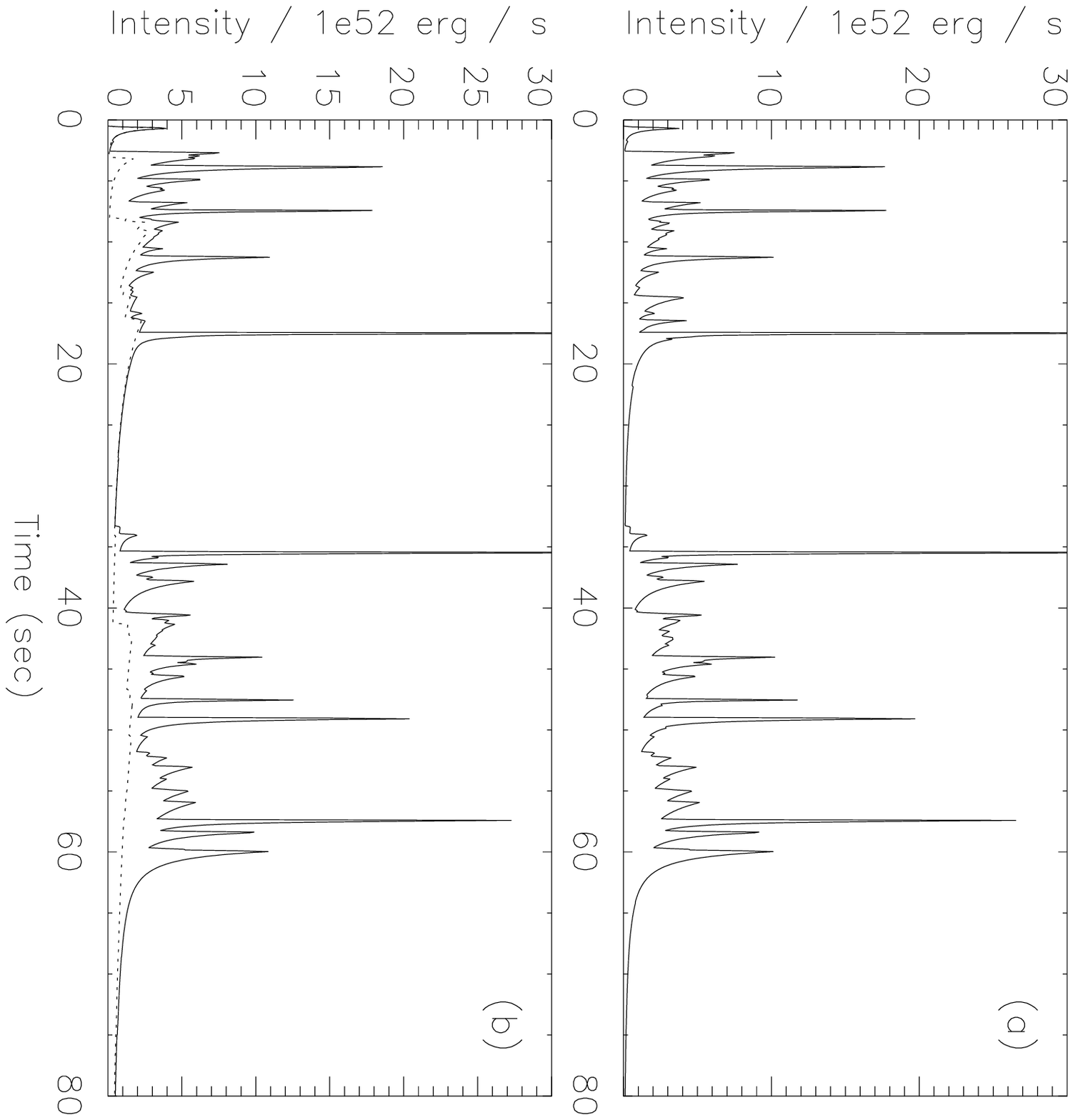]{
Simulated GRB time histories.
(a) A simulated GRB time history from internal shocks with no
deceleration.
The Lorentz factors were chosen uniformly between $10^2$ and $10^{4.5}$
and
the peak luminosity  (found over 256 ms) is $L_{256} = 2 \times 10^{53}$ erg
s$^{-1}$. Shells were randomly created at the central site for 60 sec
except for a 13 s period, thus producing a gap in the simulated GRB time
history. About 25\% of the bulk motion was converted into radiation by the
internal shocks.
(b) Simulated GRB time history from internal shocks including
deceleration of the first shell due to an ISM with 1 particle cm$^{-3}$.
The
same random numbers as in (a) were used. The lower curve is the
contribution to the time history from collisions with the decelerating
shell. Note that the there are still many narrow spikes throughout the
event. There is a slower decay during the gap. About 45\% of the bulk
motion was converted into radiation by the
internal shocks.
\label{thfig}
}
\figcaption[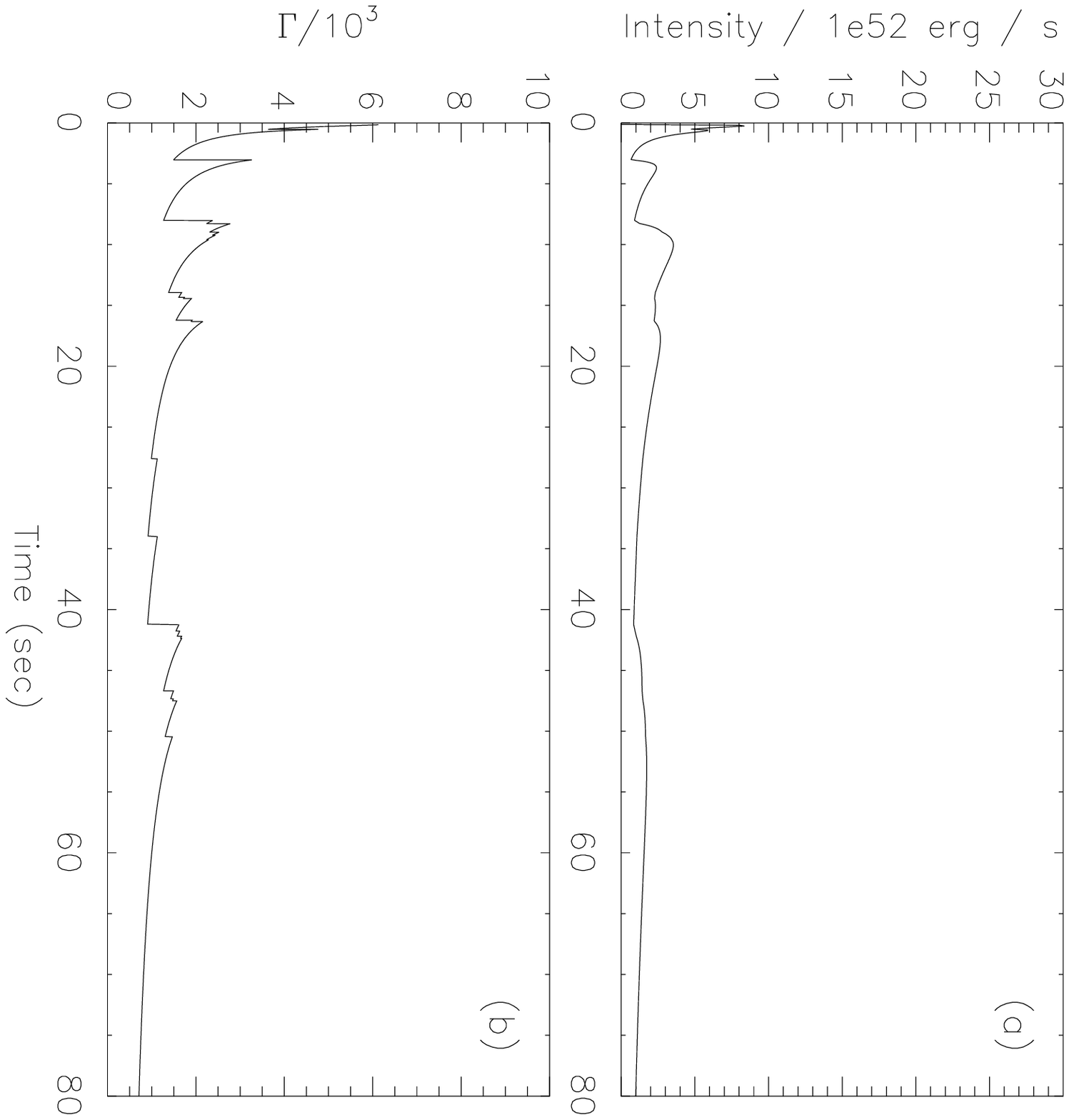]{
(a) The Lorentz factor of the decelerating shell in Figure \ref{thfig} as
a function
of when photons would arrive from it.  The Lorentz factor is fairly
constant for much of the burst because shells continue to collide with it.
(b) The expected time history from the bulk motion energy that is
converted into internal energy by the collision of the first shell with
the ISM. It is smooth and would fill in gaps in the time history.
\label{th_gamma}
}
\figcaption[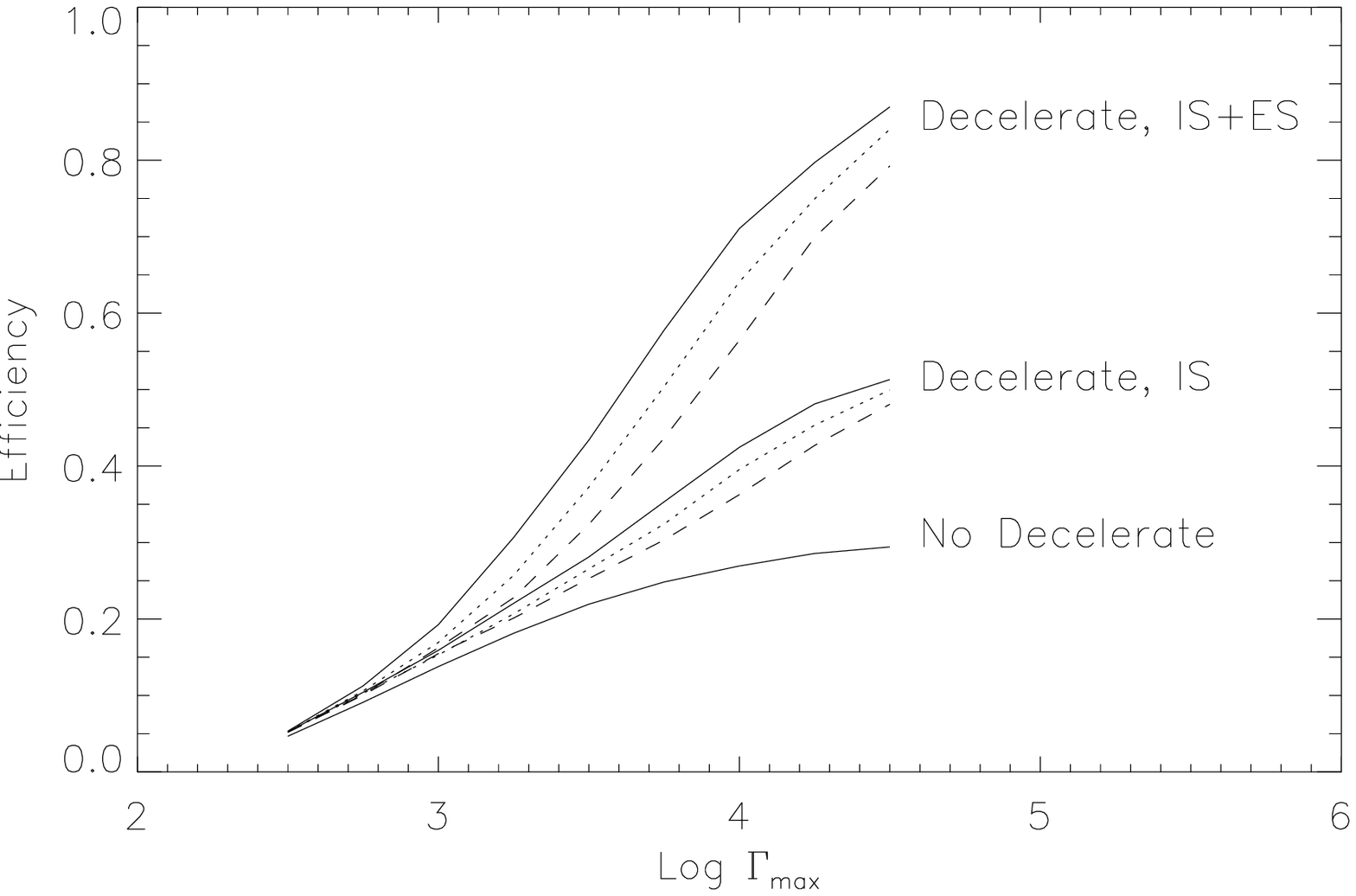]{
The efficiency of converting bulk motion into internal energy as a
function of the maximum Lorentz factor.  If there is no deceleration (ISM
density = 0), the maximum efficiency is about 25\%. If deceleration is
included, the internal shocks convert up to 45\% of the bulk motion to
internal energy. The deceleration is caused by an external shock that
sweeps up and energizes the ISM. The curve labeled ``Deceleration, IS+ES"
gives the fraction of the total original bulk energy that is lost from the
shells (some into internal shocks, some to energized the ISM). Up to 85\%
of the original energy is used during the GRB phase if the Lorentz factor
is as large as $3 \times 10^4$.
\label{eff_l}
}
\figcaption[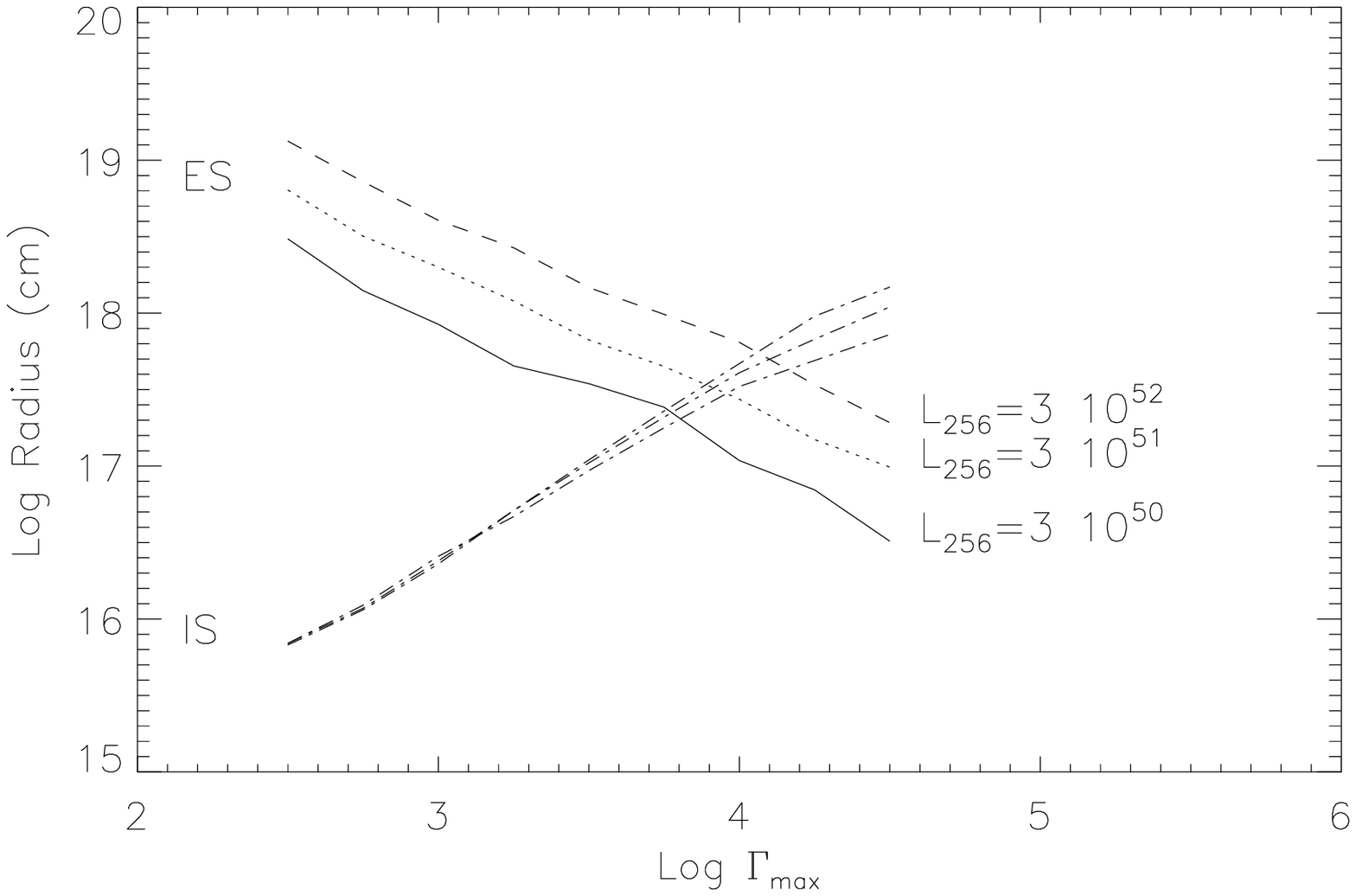]{
The radii for deceleration and internal shocks.  The solid, dotted, and
dashed lines are for increasing values of $L_{256}$. 
The curves labeled ``ES'' are the average radii when the Lorentz factor of
the first shell first drops to half its original value.  The curves
labeled ``IS'' are the average radii for  internal shocks that produce
peaks during the GRB phase.
For Lorentz factors
greater than $\sim 3000$, many of the internal shocks are caused by shells
running into the back of the first shell which is undergoing deceleration.
\label{radius_l}
}
\figcaption[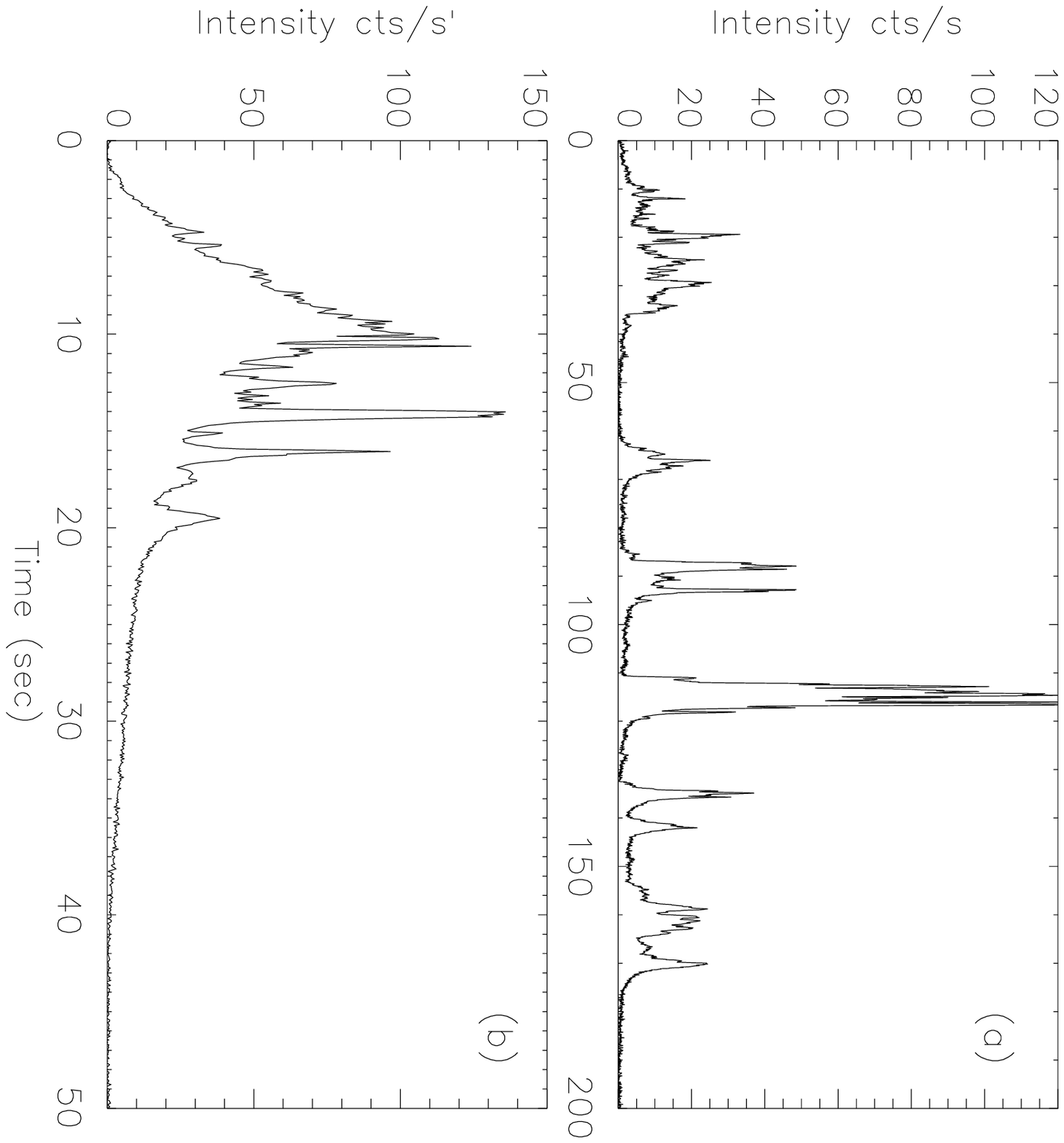]{
The time history of  long BATSE GRBs. (a)BATSE burst 2831 which is
consistent with internal shocks with very little deceleration. Note that
the gaps are near background and the decay into the gaps is sharp.
(b)BATSE burst 2329 which is consistent with substantial deceleration.
Note that the peaks tend to get progressively wider and there is an
underlying smooth component.
\label{batseth}
}



\clearpage

\centerline{Figure 1}
\plotone{thfig.eps}
\clearpage
\centerline{Figure 2}
\plotone{th_gamma.eps}
\clearpage
\centerline{Figure 3}
\plotone{eff_l.eps}
\clearpage
\centerline{Figure 4}
\plotone{radius_l.eps}
\clearpage
\centerline{Figure 5}
\plotone{batseth.eps}

\end{document}